\documentclass[twocolumn,showpacs,preprintnumbers,amsmath,amssymb]{revtex4}
\usepackage{graphicx}
\usepackage{dcolumn}
\usepackage{bm}

\begin{document}

\preprint{APS/123-QED}

\title{Comment on ``Quantum key distribution via quantum encryption'' [Phys. Rev. A 64, 024302 (2001)]}

\author{Fei Gao$^{1,2,}$\footnote{Electronic address: hzpe@sohu.com}, Sujuan Qin$^{1}$, Qiaoyan Wen$^{1}$, and Fuchen Zhu$^{3}$}
\affiliation{
$^{1}$School of Science, Beijing University of Posts and Telecommunications, Beijing, 100876, China\\
$^{2}$State Key Laboratory of Integrated Services Network, Xidian
University, Xi'an, 710071, China\\
$^{3}$National Laboratory for Modern Communications, P.O.Box 810,
Chengdu, 610041, China}
\date{\today}

\begin{abstract}
In the paper [Zhang, Li and Guo, Phys. Rev. A \textbf{64}, 024302
(2001)], a quantum key distribution protocol based on quantum
encryption was proposed, in which the quantum key can be reused.
However, it is shown that, if Eve employs a special strategy to
attack, this protocol becomes insecure because of the reused
quantum key. That is, Eve can elicit partial information about the
key bits without being detected. Finally, a possible improvement
of the Zhang-Li-Guo protocol is proposed.
\end{abstract}


\maketitle

In Ref.\cite {ZLG}, Zhang, Li and Guo proposed a quantum key
distribution protocol based on quantum encryption. This protocol
employs previously shared EPR pairs as a quantum key to encode and
decode the classical cryptography key, and the quantum key is
reusable. However, here we will show that, this protocol would
become insecure if the quantum key is reused for more than two
times.

\begin{figure}
\includegraphics[width=3.3in]{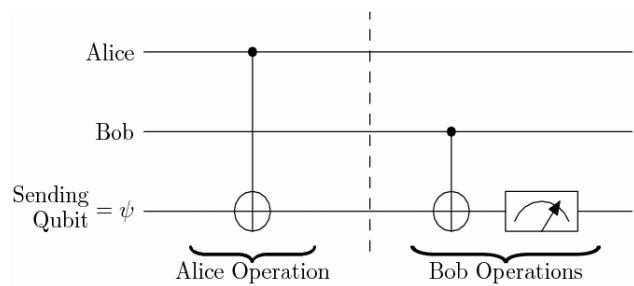}
\caption{\label{fig:one} The Zhang-Li-Guo protocol. Note that in
this Comment, for simplicity, the operation $R(\pi/4)\otimes
R(\pi/4)$ or $R(\pi/4)^{\otimes 3}$ is not included in our
figures.}
\end{figure}

For convenience, except for especial declarations, we use the same
notations as in Ref.\cite {ZLG}. Let us give a brief description
of the Zhang-Li-Guo protocol firstly (see Fig.~\ref{fig:one}). At
the beginning, Alice and Bob share some quantity of EPR pairs
serving as the quantum key:
$|\Phi^+\rangle=1/\sqrt{2}(|00\rangle+|11\rangle)$. To send the
key bit (0 or 1) to Bob, Alice prepares a carrier particle
$\gamma$ in the corresponding state $|\psi\rangle$ ($|0\rangle$ or
$|1\rangle$), performs a controlled-{\footnotesize NOT}
({\footnotesize CNOT}) operation on $\gamma$ and thus entangles
this qubit to the previously shared Bell state. Then she transmits
this qubit to Bob, from which Bob can obtain the key bit $\psi$ by
performing a {\footnotesize CNOT} operation and a measurement on
it. Because every sending qubit is in a completely mixed state,
Eve can not extract information about the key bit. Furthermore, to
strengthen the security of this protocol, Alice and Bob perform a
rotation
\begin{eqnarray}
R(\frac{\pi}{4})=\frac{1}{\sqrt{2}} \left( \begin{array}{c c} 1 & 1 \\
-1 & 1
\end{array} \right)
\end{eqnarray}
on their respective shared particles before encrypting each
$|\psi\rangle$.

It is well known that the shared particles in Bell state have
strong quantum correlation (i.e., entanglement). It is this
correlation that makes the quantum encryption secure. The author
of Ref.\cite{ZLG} argues that, because this correlation cannot be
produced by LQCC and the eavesdropper cannot establish this
correlation with the sender, the quantum key is reusable. However,
they overlooked a fact that the sending qubit would bring Eve the
chance to entangle her ancilla to the shared Bell state, which
means that the eavesdropper can establish this correlation with
the sender. As a result, this protocol becomes insecure when the
quantum key is reused.

\begin{figure}
\includegraphics[width=3.3in]{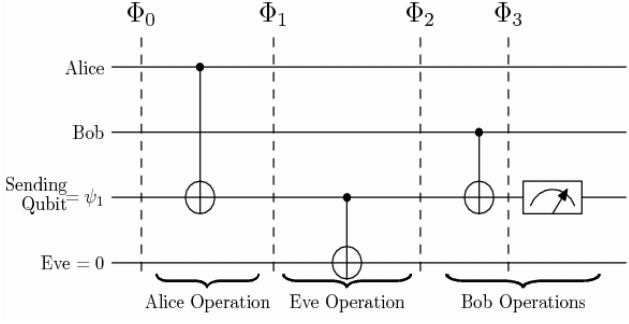}
\caption{\label{fig:two} Eve's attack in the first round.}
\end{figure}

Now we come to Eve's eavesdropping strategy. Consider a certain
EPR pair shared by Alice and Bob, which will be used to encrypt
$\gamma_1, \gamma_2, \gamma_3,...$ (the corresponding states are
$|\psi_1\rangle, |\psi_2\rangle, |\psi_3\rangle,...$ respectively,
where $\psi_i=0$ or $1$). Hereafter we use the term ``the $i$-th
round" to denote the processing procedures of $\gamma_i$, and
Alice and Bob's operation $R(\pi/4)\otimes R(\pi/4)$ is taken as
the beginning of each round. Furthermore, we use
$|\phi_{i0}\rangle_{A,B,E}$ and $|\phi_{i1}\rangle_{A,B,E}$ to
denote the states shared by Alice, Bob and Eve at the beginning
and the end of the $i$-th round, respectively. In addition, the
subscriptions A, B and E represent the particles belong to Alice,
Bob, and Eve respectively, and $\gamma$ represents the sending
particle. Suppose Eve prepares $|0\rangle$ as her ancilla, the
eavesdropping strategy can be described as follows:

(i) In the first round, Eve entangles her ancilla into the Bell
state shared by Alice and Bob. More specifically, Eve intercepts
the sending qubit and performs a {\footnotesize CNOT} operation on
her ancilla, then resends the sending qubit to Bob (see
Fig.~\ref{fig:two}). The initial state of Alice, Bob and Eve's
particles can be represented as
\begin{eqnarray}
|\phi_{10}\rangle_{A,B,E}=\frac{1}{\sqrt{2}}(|0,0,0\rangle+|1,1,0\rangle)_{A,B,E}.
\end{eqnarray}
Then the states at various stages in Fig.~\ref{fig:two} are as
follows:
\begin{eqnarray}
|\Phi_0\rangle&=&\frac{1}{\sqrt{2}}(|0,0,\psi_1,0\rangle+|1,1,\psi_1,0\rangle)_{A,B,\gamma,E},\\
|\Phi_1\rangle&=&\frac{1}{\sqrt{2}}(|0,0,\psi_1,0\rangle+|1,1,\overline{\psi}_1,0\rangle)_{A,B,\gamma,E},\\
|\Phi_2\rangle&=&\frac{1}{\sqrt{2}}(|0,0,\psi_1,\psi_1\rangle+|1,1,\overline{\psi}_1,\overline{\psi}_1\rangle)_{A,B,\gamma,E},\\
|\Phi_3\rangle&=&\frac{1}{\sqrt{2}}(|0,0,\psi_1,\psi_1\rangle+|1,1,\psi_1,\overline{\psi}_1\rangle)_{A,B,\gamma,E},
\end{eqnarray}
where the overline expresses bit flip, for example,
$\overline{\psi}_1=\psi_1+1$ modulo 2.

In the last stage, when Bob performs his {\footnotesize CNOT}
operation, he disentangles the sending qubit $|\psi_1\rangle$ and
correctly gets the value of $\psi_1$, while the original Bell
state has now been entangled with the state of Eve in the form of
\begin{equation}
|\phi_{11}\rangle_{A,B,E}=\frac{1}{\sqrt{2}}(|0,0,\psi_1\rangle+|1,1,\overline{\psi}_1\rangle)_{A,B,E}.
\end{equation}

(ii) In the second round, Eve tries to avoid the detection and, at
the same time, retain her entanglement with Alice and Bob. As was
proved in Ref.\cite{ZLG}, Eve can not obtain information in this
round. However, we will show that she can take some measures to
avoid the detection.

Firstly, when Alice and Bob perform the operations
$R(\pi/4)\otimes R(\pi/4)$ on their ``Bell state", Eve also
performs $R(\pi/4)$ on her ancilla. As a result, the entangled
state of Alice, Bob and Eve will be converted into
\begin{eqnarray}
|\phi_{20}\rangle_{A,B,E}&=&R(\frac{\pi}{4})^{\otimes 3} |\phi_{11}\rangle_{A,B,E}\nonumber\\
&=&\frac{1}{2}\big[|0,0,0\rangle+(-1)^{\psi_1}|0,1,1\rangle\nonumber\\
& &+(-1)^{\psi_1}|1,0,1\rangle+|1,1,0\rangle \big]_{A,B,E},
\end{eqnarray}
where the identity
$R(\frac{\pi}{4})|\psi\rangle=1/\sqrt{2}\big[|0\rangle+(-1)^{\overline{\psi}}|1\rangle\big]$
was used.
\begin{figure}
\includegraphics[width=3.3in]{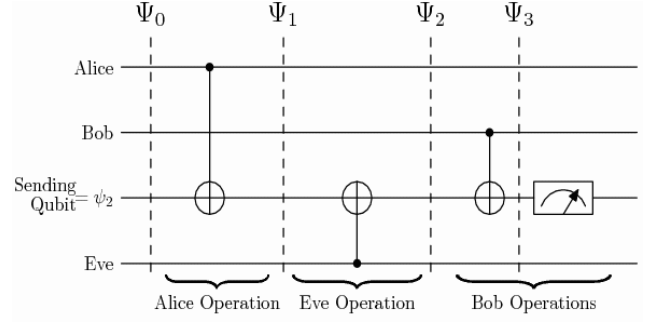}
\caption{\label{fig:three} Eve's attack in the second round.}
\end{figure}

Afterwards, Eve intercepts the sending qubit, performs a
{\footnotesize CNOT} operation on it, and then resends it to Bob
(see Fig.~\ref{fig:three}). The states at various stages in
Fig.~\ref{fig:three} are as follows:
\begin{eqnarray}
|\Psi_0\rangle&=&\frac{1}{2}\big[|0,0,\psi_2,0\rangle+(-1)^{\psi_1}|0,1,\psi_2,1\rangle\nonumber\\
&+&(-1)^{\psi_1}|1,0,\psi_2,1\rangle+|1,1,\psi_2,0\rangle\big]_{A,B,\gamma,E},\\
|\Psi_1\rangle&=&\frac{1}{2}\big[|0,0,\psi_2,0\rangle+(-1)^{\psi_1}|0,1,\psi_2,1\rangle\nonumber\\
&+&(-1)^{\psi_1}|1,0,\overline{\psi}_2,1\rangle+|1,1,\overline{\psi}_2,0\rangle\big]_{A,B,\gamma,E},\\
|\Psi_2\rangle&=&\frac{1}{2}\big[|0,0,\psi_2,0\rangle+(-1)^{\psi_1}|0,1,\overline{\psi}_2,1\rangle\nonumber\\
&+&(-1)^{\psi_1}|1,0,\psi_2,1\rangle+|1,1,\overline{\psi}_2,0\rangle\big]_{A,B,\gamma,E},\\
|\Psi_3\rangle&=&\frac{1}{2}\big[|0,0,\psi_2,0\rangle+(-1)^{\psi_1}|0,1,\psi_2,1\rangle\nonumber\\
&+&(-1)^{\psi_1}|1,0,\psi_2,1\rangle+|1,1,\psi_2,0\rangle\big]_{A,B,\gamma,E}.
\end{eqnarray}

In the last stage, when Bob performs his {\footnotesize CNOT}
operation, he disentangles the sending qubit $|\psi_2\rangle$ and
correctly gets the value of $\psi_2$, while leaving the state
\begin{eqnarray}
|\phi_{21}\rangle_{A,B,E}&=&\frac{1}{2}\big[|0,0,0\rangle+(-1)^{\psi_1}|0,1,1\rangle\nonumber\\
& &+(-1)^{\psi_1}|1,0,1\rangle+|1,1,0\rangle \big]_{A,B,E}.
\end{eqnarray}

(iii) In the third round, Eve eavesdrops the key bit. Firstly, as
in step (ii), Eve also performs $R(\pi/4)$ on her ancilla when
Alice and Bob perform $R(\pi/4)$ on their respective particles.
The entangled state will be changed into
\begin{eqnarray}
|\phi_{30}\rangle_{A,B,E}&=&R(\frac{\pi}{4})^{\otimes 3} |\phi_{21}\rangle_{A,B,E}\nonumber\\
&=&\frac{1}{2\sqrt{2}}\big[\alpha\left(|0,0,0\rangle-|1,1,1\rangle\right)\nonumber\\
& &-\beta\left(|0,0,1\rangle-|1,1,0\rangle\right)\big]_{A,B,E},
\end{eqnarray}
where $\alpha=1+(-1)^{\psi_1}$, $\beta=1-(-1)^{\psi_1}$.
\begin{figure}
\includegraphics[width=3.3in]{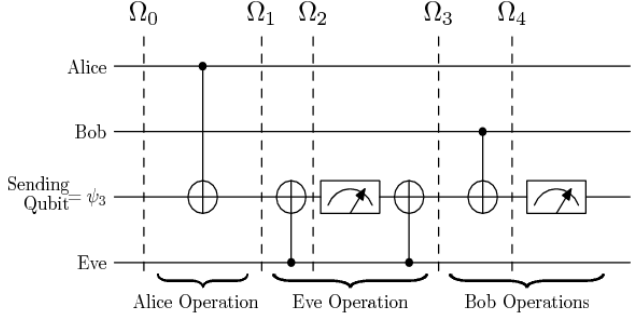}
\caption{\label{fig:four} Eve's attack in the third round.}
\end{figure}

Afterwards, Eve intercepts the sending qubit, performs a
{\footnotesize CNOT} operation, a measurement and another
{\footnotesize CNOT} operation on it, and then resends it to Bob
(see Fig.~\ref{fig:four}). The states at various stages in
Fig.~\ref{fig:four} are as follows:
\begin{eqnarray}
|\Omega_0\rangle&=&\frac{1}{2\sqrt{2}}\big[\alpha\left(|0,0,\psi_3,0\rangle-|1,1,\psi_3,1\rangle\right)\nonumber\\
&-&\beta\left(|0,0,\psi_3,1\rangle-|1,1,\psi_3,0\rangle\right)\big]_{A,B,\gamma,E},\\
|\Omega_1\rangle&=&\frac{1}{2\sqrt{2}}\big[\alpha\left(|0,0,\psi_3,0\rangle-|1,1,\overline{\psi}_3,1\rangle\right)\nonumber\\
&-&\beta\left(|0,0,\psi_3,1\rangle-|1,1,\overline{\psi}_3,0\rangle\right)\big]_{A,B,\gamma,E},\\
|\Omega_2\rangle&=&\frac{1}{2\sqrt{2}}\big[\alpha\left(|0,0,\psi_3,0\rangle-|1,1,\psi_3,1\rangle\right)\nonumber\\
&-&\beta\left(|0,0,\overline{\psi}_3,1\rangle-|1,1,\overline{\psi}_3,0\rangle\right)\big]_{A,B,\gamma,E},\\
|\Omega_3\rangle&=&\frac{1}{2\sqrt{2}}\big[\alpha\left(|0,0,\psi_3,0\rangle-|1,1,\overline{\psi}_3,1\rangle\right)\nonumber\\
&-&\beta\left(|0,0,\psi_3,1\rangle-|1,1,\overline{\psi}_3,0\rangle\right)\big]_{A,B,\gamma,E},\\
|\Omega_4\rangle&=&\frac{1}{2\sqrt{2}}\big[\alpha\left(|0,0,\psi_3,0\rangle-|1,1,\psi_3,1\rangle\right)\nonumber\\
&-&\beta\left(|0,0,\psi_3,1\rangle-|1,1,\psi_3,0\rangle\right)\big]_{A,B,\gamma,E}.
\end{eqnarray}

It can be seen that Eve disentangles the key qubit by a
{\footnotesize CNOT} operation, and then restores the entangled
state by another {\footnotesize CNOT} operation after a
measurement. As a result, Eve obtains the measurement result
$\psi_3+\psi_1$ (modulo 2) and Bob correctly gets the value of
$\psi_3$. At last, the entangled state of Alice, Bob and Eve can
be written as
\begin{eqnarray}
|\phi_{31}\rangle_{A,B,E}&=&\frac{1}{2\sqrt{2}}\big[\alpha\left(|0,0,0\rangle-|1,1,1\rangle\right)\nonumber\\
& &-\beta\left(|0,0,1\rangle-|1,1,0\rangle\right)\big]_{A,B,E}.
\end{eqnarray}

\begin{figure}
\includegraphics[width=3.3in]{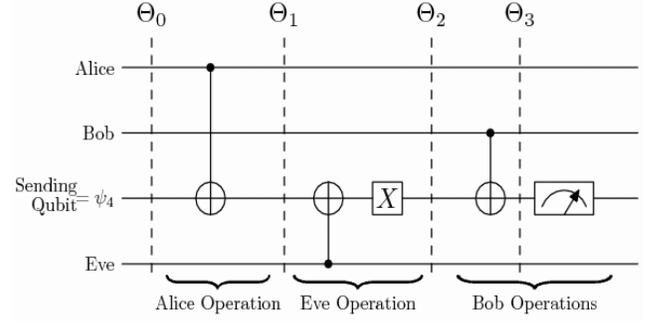}
\caption{\label{fig:five} Eve's attack in the fourth round.}
\end{figure}

(iv) In the fourth round, Eve uses a similar strategy as in the
second round to avoid the detection, the only difference is that
Eve has to perform an additional $X=\left(\begin{array}{l l}0, & 1
\\ 1, & 0 \end{array}\right)$ operation on the sending qubit here (see
Fig.~\ref{fig:five}). After their operation
$R(\frac{\pi}{4})^{\otimes 3}$, Alice, Bob and Eve change the
entangled state into
\begin{eqnarray}
|\phi_{40}\rangle_{A,B,E}&=&R(\frac{\pi}{4})^{\otimes 3}|\phi_{31}\rangle_{A,B,E}\nonumber\\
&=&-\frac{1}{2}\big[|0,0,1\rangle+(-1)^{\psi_1}|0,1,0\rangle\nonumber\\
& &+(-1)^{\psi_1}|1,0,0\rangle+|1,1,1\rangle \big]_{A,B,E}.
\end{eqnarray}

Then Eve performs the operations as described in
Fig.~\ref{fig:five}. The states at various stages are as follows:
\begin{eqnarray}
|\Theta_0\rangle&=&-\frac{1}{2}\big[|0,0,\psi_4,1\rangle+(-1)^{\psi_1}|0,1,\psi_4,0\rangle\nonumber\\
&+&(-1)^{\psi_1}|1,0,\psi_4,0\rangle+|1,1,\psi_4,1\rangle\big]_{A,B,\gamma,E},\\
|\Theta_1\rangle&=&-\frac{1}{2}\big[|0,0,\psi_4,1\rangle+(-1)^{\psi_1}|0,1,\psi_4,0\rangle\nonumber\\
&+&(-1)^{\psi_1}|1,0,\overline{\psi}_4,0\rangle+|1,1,\overline{\psi}_4,1\rangle\big]_{A,B,\gamma,E},\\
|\Theta_2\rangle&=&-\frac{1}{2}\big[|0,0,\psi_4,1\rangle+(-1)^{\psi_1}|0,1,\overline{\psi}_4,0\rangle\nonumber\\
&+&(-1)^{\psi_1}|1,0,\psi_4,0\rangle+|1,1,\overline{\psi}_4,1\rangle\big]_{A,B,\gamma,E},\\
|\Theta_3\rangle&=&-\frac{1}{2}\big[|0,0,\psi_4,1\rangle+(-1)^{\psi_1}|0,1,\psi_4,0\rangle\nonumber\\
&+&(-1)^{\psi_1}|1,0,\psi_4,0\rangle+|1,1,\psi_4,1\rangle\big]_{A,B,\gamma,E}.
\end{eqnarray}

It can be seen that, in the last stage, Bob correctly gets the
value of $\psi_4$, while leaving the state
\begin{eqnarray}
|\phi_{41}\rangle_{A,B,E}&=&-\frac{1}{2}\big[|0,0,1\rangle+(-1)^{\psi_1}|0,1,0\rangle\nonumber\\
& &+(-1)^{\psi_1}|1,0,0\rangle+|1,1,1\rangle \big]_{A,B,E}.
\end{eqnarray}

(v) In the fifth round, Eve uses the same strategy as in the third
round to eavesdrop the key bit, that is, the strategy in step
(iii). After their operation $R(\frac{\pi}{4})^{\otimes 3}$,
Alice, Bob and Eve change the entangled state into
\begin{eqnarray}
|\phi_{50}\rangle_{A,B,E}&=&R(\frac{\pi}{4})^{\otimes 3}|\phi_{41}\rangle_{A,B,E}\nonumber\\
&=&-\frac{1}{2\sqrt{2}}\big[\alpha(|0,0,0\rangle+|1,1,1\rangle)\nonumber\\
& &+\beta(|0,0,1\rangle+|1,1,0\rangle)\big]_{A,B,E}.
\end{eqnarray}

Then Eve performs the operations as described in
Fig.~\ref{fig:four}. The states at various stages are as follows:
\begin{eqnarray}
|\Upsilon_0\rangle&=&-\frac{1}{2\sqrt{2}}\big[\alpha\left(|0,0,\psi_5,0\rangle+|1,1,\psi_5,1\rangle\right)\nonumber\\
& &+\beta\left(|0,0,\psi_5,1\rangle+|1,1,\psi_5,0\rangle\right)\big]_{A,B,\gamma,E},\\
|\Upsilon_1\rangle&=&-\frac{1}{2\sqrt{2}}\big[\alpha\left(|0,0,\psi_5,0\rangle+|1,1,\overline{\psi}_5,1\rangle\right)\nonumber\\
& &+\beta\left(|0,0,\psi_5,1\rangle+|1,1,\overline{\psi}_5,0\rangle\right)\big]_{A,B,\gamma,E},
\end{eqnarray}
\begin{eqnarray}
|\Upsilon_2\rangle&=&-\frac{1}{2\sqrt{2}}\big[\alpha\left(|0,0,\psi_5,0\rangle+|1,1,\psi_5,1\rangle\right)\nonumber\\
& &+\beta\left(|0,0,\overline{\psi}_5,1\rangle+|1,1,\overline{\psi}_5,0\rangle\right)\big]_{A,B,\gamma,E},\\
|\Upsilon_3\rangle&=&-\frac{1}{2\sqrt{2}}\big[\alpha\left(|0,0,\psi_5,0\rangle+|1,1,\overline{\psi}_5,1\rangle\right)\nonumber\\
& &+\beta\left(|0,0,\psi_5,1\rangle+|1,1,\overline{\psi}_5,0\rangle\right)\big]_{A,B,\gamma,E},\\
|\Upsilon_4\rangle&=&-\frac{1}{2\sqrt{2}}\big[\alpha\left(|0,0,\psi_5,0\rangle+|1,1,\psi_5,1\rangle\right)\nonumber\\
& &+\beta\left(|0,0,\psi_5,1\rangle+|1,1,\psi_5,0\rangle\right)\big]_{A,B,\gamma,E},
\end{eqnarray}
where $\Upsilon_p$ corresponds to the state $\Omega_p$ in
Fig.~\ref{fig:four} ($p=0,1,2,3,4$). It can be seen that Eve's
measurement result in this round is $\psi_5+\psi_1$ (modulo 2).

Obviously, in the last stage, Bob correctly gets the value of
$\psi_5$, while leaving the state
\begin{eqnarray}
|\phi_{51}\rangle_{A,B,E}&=&-\frac{1}{2\sqrt{2}}\big[\alpha(|0,0,0\rangle+|1,1,1\rangle)\nonumber\\
& &+\beta(|0,0,1\rangle+|1,1,0\rangle)\big]_{A,B,E}.
\end{eqnarray}

Comparing the state $|\phi_{51}\rangle_{A,B,E}$ with
$|\phi_{11}\rangle_{A,B,E}$, we can verify that the two states is
equivalent except for a global phase factor (i.e., $-1$). That is,
from an observational point of view these two states are identical
\cite{QCQI}. Therefore, in the following rounds, Eve can use the
same strategy as in the steps from (ii) to (v) repeatedly.

Now let us give a concretely description of our eavesdropping
strategy:
\begin{description}
\item [\quad 1.] In the first round, Eve performs the operations
as described in Fig.~\ref{fig:two}; \item [\quad 2.] When Alice
and Bob perform $R(\frac{\pi}{4})$ on their respective particles
at the beginning of every round (except for the first round), Eve
also performs $R(\frac{\pi}{4})$ on her ancilla; \item [\quad 3.]
From the second round to the fifth round, Eve performs the
operations as described in Fig.~\ref{fig:three},
Fig.~\ref{fig:four}, Fig.~\ref{fig:five} and Fig.~\ref{fig:four}
in turn; \item [\quad 4.] In the following rounds, Eve performs
the operations as described in item $3$ repeatedly.
\end{description}

From the above analysis, we can see that in our eavesdropping
strategy no error will be introduced to the key distribution
between Alice and Bob, and Eve will obtain exactly the result of
$$\psi_3+\psi_1,\psi_5+\psi_1,\psi_7+\psi_1,\psi_9+\psi_1,\dots$$
from which she can infer about half of the key bits by checking
two possible values for $\psi_1$. It should be emphasized that
there is another profitable fact for Eve. That is, at the end of
QKD procedure, Alice and Bob will compare a subsequence of the key
bits publicly to detect eavesdropping, which obviously leak useful
information to Eve.

Now it is worthwhile to inspect the basic idea of our attack
strategy. Though Eve cannot get information about the key bit in
every even rounds (as proved in Ref.\cite{ZLG}), she can take some
more clever measures to avoid the detection and retain her
entanglement with Alice and Bob so that she can eavesdrop the key
bit in the next round. Our attack strategy is exactly based on
this fact. By our strategy, if the shared Bell states are reused
for many times, Eve can obtain about half of the key bits without
being detected by Alice and Bob. One may argue that the shared
Bell states would not be reused for too many times without special
treatments by Alice and Bob, such as quantum privacy amplification
and entanglement purification \cite{ZLG}. However, from above
analysis it can be seen that Eve needs only three rounds to elicit
partial information about the key bits, which definitely forms a
serous threaten to the Zhang-Li-Guo protocol. In fact, the QKD
protocols in Refs.\cite{KBB,BK} have similar hidden troubles, see
Refs.\cite{GGWZ1,GGWZ2} for details.

Before we conclude, let us give a discussion about the rotation
\begin{eqnarray}
R(\theta)=\left( \begin{array}{c c} \cos\theta & \sin\theta \\
-\sin\theta & \cos\theta
\end{array} \right),
\end{eqnarray}
which plays an important role in the Zhang-Li-Guo protocol.
Without Alice and Bob's rotations at the beginning of every round,
this QKD protocol would be insecure. For example, in this
condition Eve can entangle her ancilla into the Bell state in the
first round (as described in Fig.~\ref{fig:two}), and then elicit
information about the key bits in the following rounds (as
described in Fig.~\ref{fig:four}). As a result, Eve will obtain
the result of
$$\psi_2+\psi_1,\psi_3+\psi_1,\psi_4+\psi_1,\psi_5+\psi_1,\dots.$$
(To avoid confusion we call this attack strategy $S_1$, and call
the strategy we showed in above paragraphs $S_2$.) Therefore, the
rotations are necessary, and $\pi/4$ is selected as the rotation
angle because it leads to the maximum error rate (i.e., $1/2$)
caused by Eve when she uses the strategy $S_1$ \cite{ZLG}.
However, it is the selection of $\theta=\pi/4$ that makes the
Zhang-Li-Guo protocol insecure against $S_2$. That is, the error
rate caused by Eve is $0$ when she uses the strategy $S_2$.
Hereafter we use $d_1$ and $d_2$ to denote the error rate
corresponding to $S_1$ and $S_2$, respectively. In fact, it is not
difficult to prove that, if $\theta\neq k\pi\pm\pi/4$
($k=0,\pm1,\pm2,...$), it is impossible for Eve to elicit
information about the key bits without introducing disturbance
(See the Appendix for details). Consequently, by altering
$\theta$, we can modify the Zhang-Li-Guo protocol so that it can
resist both $S_1$ and $S_2$.

As was given in Ref.\cite{ZLG}, when Eve uses $S_1$ to attack, the
error rate is $d_1=2\cos^2\theta\sin^2\theta$. By similar
deduction we can obtain the error rate when $S_2$ is used, i.e.,
$d_2=\frac{1}{2}(\sin^2\theta-\cos^2\theta)^2$. Clearly, there is
a trade-off between $d_1$ and $d_2$, which satisfy the ralation of
$d_1+d_2=1/2$. That is, a greater $d_1$ results in a smaller
$d_2$, and vice versa. It can be seen that $\theta=\pi/4$ is a
extreme instance, where $d_1$ reaches its maximum value $1/2$ but
$d_2=0$. Therefore, we can select such a rotation angle (denoted
as $\theta_0$) that $d_1=d_2=1/4$, i.e.,
$2\cos^2\theta_0\sin^2\theta_0=1/4$. As a result, when we use
$\theta_0$ instead of $\pi/4$ in the Zhang-Li-Guo protocol, it can
resist both attack strategies (because either strategy will
introduce an error rate of $1/4$). We have to confess that this
modification decreases the efficiency of eavesdropping detection.
However, $1/4$ is still a sufficient value for a detection
probability. In fact, as far as the general intercept-resend
strategy is concerned, the detection probability in BB84 protocol
\cite{BB84} is $1/4$, too.

In summary, we have presented a special attack strategy to the
Zhang-Li-Guo protocol \cite{ZLG}, in which Eve can elicit partial
information about the key bits without being detected when the
quantum key is reused for more than two times. Furthermore, we
have discussed about the ralation between the security and the
value of $\theta$, and pointed out that this QKD protocol would be
secure if we use $\theta_0$ instead of $\pi/4$.

This work is supported by the National Natural Science Foundation
of China, Grants No. 60373059; the National Laboratory for Modern
Communications Science Foundation of China, Grants No.
51436020103DZ4001; the National Research Foundation for the
Doctoral Program of Higher Education of China, Grants No.
20040013007; and the ISN Open Foundation.

\appendix*
\section{}
In this appendix we will show that when $\theta\neq k\pi\pm\pi/4$
($k=0,\pm1,\pm2,...$), it is inevitable for Eve to introduce
disturbance if she has entangled her ancilla into the Bell state
in the first round.

Without loss of generality, suppose that in the first round Eve's
system has entangled with Alice and Bob's key in the state
\begin{eqnarray}
|\Lambda\rangle=\frac{1}{\sqrt{2}}(|00\rangle|\varphi_0\rangle+|11\rangle|\varphi_1\rangle)_{A,B,E},\label{eq:Ac}
\end{eqnarray}
where there is no restriction on the form of $|\varphi_0\rangle$
and $|\varphi_1\rangle$. After Alice and Bob do a bilateral
rotation $R(\theta)$, Alice does a {\footnotesize CNOT} operation
on the sending qubit $|\psi_2\rangle$ and sends it out. Then Eve
does a unitary transformation on the sending qubit and her own
system. She expects that Alice and Bob cannot detect her existence
(i.e., the error rate caused by her is $0$). Assume that the
unitary transformation has the universal form
\begin{eqnarray}
U_{\gamma,E}|i\rangle_{\gamma}|\varphi_j\rangle_E=(a_{ij}|0\rangle|\varphi_{aij}\rangle+b_{ij}|1\rangle|\varphi_{bij}\rangle)_{\gamma,E},
\end{eqnarray}
where $i,j=0,1$ and there is no restriction on the final state of
$|\varphi\rangle_E$. At last, Bob receives the sending qubit and
uses a {\footnotesize CNOT} operation to disentangle it from the
shared state.

Suppose that the composite system
$|\Lambda\rangle_{A,B,E}\otimes|\psi_2\rangle_{\gamma}$ is changed
into $|\Delta\rangle$ after all the above operations, we can
easily write the form of the state $|\Delta\rangle$. If the attack
is successful, it requires that the sending qubit $|\psi_2\rangle$
is correctly disentangled by Bob. To satisfy this requirement, we
obtain the following results:

When $\psi_2=0$, we get
\begin{eqnarray}
b_{00}\cos^2\theta|\varphi_{b00}\rangle+b_{01}\sin^2\theta|\varphi_{b01}\rangle=0,\\
-a_{00}\sin\theta\cos\theta|\varphi_{a00}\rangle+a_{01}\sin\theta\cos\theta|\varphi_{a01}\rangle=0,\label{eq:Aa}\\
-b_{10}\sin\theta\cos\theta|\varphi_{b10}\rangle+b_{11}\sin\theta\cos\theta|\varphi_{b11}\rangle=0,\label{eq:Ab}\\
a_{10}\sin^2\theta|\varphi_{a10}\rangle+a_{11}\cos^2\theta|\varphi_{a11}\rangle=0.
\end{eqnarray}
When $\psi_2=1$, we get
\begin{eqnarray}
a_{10}\cos^2\theta|\varphi_{a10}\rangle+a_{11}\sin^2\theta|\varphi_{a11}\rangle=0,\\
b_{00}\sin^2\theta|\varphi_{b00}\rangle+b_{01}\cos^2\theta|\varphi_{b01}\rangle=0,\label{eq:Ad}
\end{eqnarray}
where we omit two equations the same as Eqs.(\ref{eq:Aa}) and
(\ref{eq:Ab}).

With the help of Eqs.(\ref{eq:Ac})$\sim$(\ref{eq:Ad}), we then
obtain two possible conditions: either (1)
$|\varphi_0\rangle=|\varphi_1\rangle$, which means
$|\Lambda\rangle$ is a product state of Eve's ancilla and Alice
and Bob's Bell state; or (2) $\theta= k\pi\pm\pi/4$. This result
implies that only when $\theta= k\pi\pm\pi/4$ Eve can entangle her
ancilla into the Bell state without introducing any disturbance,
which is the exact conclusion we want to prove.

\end{document}